\documentclass[10pt]{article}
\usepackage{eurosym}
\usepackage{times}
\usepackage{hyperref}
\usepackage{graphicx}
\usepackage{epstopdf}
\usepackage{amsmath, bm}
\usepackage{amsmath}
\usepackage{amssymb}
\usepackage{physics}
\usepackage{xcolor}
\usepackage{authblk}

\usepackage{booktabs}
\usepackage{tabularx}
\usepackage{makecell}

\newcommand{\Blue}[1]{\textcolor{blue}{#1}}
\newcommand{\Red}[1]{\textcolor{black}{#1}}

\begin{document}

\title{Temporal Soliton Waves in Modern Fiber Laser Systems}
\author[1,2,4]{Shilong Liu}
\author[3]{Boris Malomed}
\affil[1]{State Key Laboratory of Precision Spectroscopy, and Hainan Institute, East China Normal University, Shanghai, 200062, China}
\affil[2]{Chongqing Key Laboratory of Precision Optics, Chongqing Institute of East China Normal University, Chongqing, 401120, China}
\affil[3]{Department of Physical Electronics, School of Electrical Engineering, Tel Aviv University, Tel Aviv, Israel}
\affil[4]{femtoQ Lab, Department of Engineering Physics, Polytechnique Montr\'{e}al, Montr\'{e}al, Qu\'{e}bec H3T 1J4, Canada}
\date{}
\maketitle

\begin{abstract}
{This chapter provides a comprehensive survey of temporal soliton-wave
 in modern fiber systems. \Blue{Temporal soliton waves constitute a broad class of nonlinear localized structures in modern fiber laser systems, arising from different balances among dispersion, nonlinearity, gain, loss, 
 spectral filtering, and various forms of cavity coupling}. We review the underlying theory, basic
experimental findings, and practical methods for the generation, real-time
characterization, and active control of soliton families, with a particular
focus on recent progress reported for mode-locked fiber-laser platforms. The
families covered in the chapter include dissipative solitons, soliton
molecules (bound states), and dispersion-managed solitons. For each species,
we focus on the pulse energy, duration, stationary and breathing states, and
emerging intelligent (machine-learning-assisted) control strategies.}

\noindent \textbf{Acronyms}

\noindent CNN convolutional neural network

\noindent CW continuous-wave (background)

\noindent DFT dispersive Fourier transform

\noindent DM dispersion management

\noindent DNN deep neural network

\noindent DS dissipative soliton

\noindent EDF erbium-doped fiber

\noindent FNLSE fractional nonlinear Schr\"{o}dinger equation

\noindent FNN feed-forward neural network

\noindent FOD fourth-order dispersion

\noindent FROG frequency-resolved optical gating

\noindent FSE (linear) fractional Schr\"{o}dinger equation

\noindent GA genetic algorithm

\noindent GDD group-delay dispersion

\noindent GVD group-velocity dispersion

\noindent LI L\'{e}vy index (the fractionality index)

\noindent LWG L\'{e}vy waveguide (a waveguide realizing fractional diffraction or dispersion)

\noindent ML machine-learning (methods)

\noindent MLFL mode-locked fiber laser

\noindent NLSE nonlinear Schr\"{o}dinger equation

\noindent P-CNN physics-embedded CNN

\noindent PINN physics-informed neural network

\noindent PQS pure-quartic soliton

\noindent RNN recurrent neural network

\noindent SA saturable absorber

\noindent SC supercontinuum

\noindent SCG supercontinuum generation

\noindent SM soliton molecule (a bound state of solitons)

\noindent SMF single-mode fiber

\noindent SPM self-phase modulation

\noindent TS time separation (between optical pulses forming a bound state,
alias SM, in the temporal domain)

\noindent XPM cross-phase modulation
\end{abstract}

\tableofcontents

\section{Introduction}

\label{sec:intro}

The modern concept of optical temporal \emph{solitons} emerged in the early
1970s with the derivation of the nonlinear Schr\"{o}dinger equation (NLSE)
as the fundamental model for pulse propagation in dispersive Kerr media.
Using the inverse-scattering transform, Zakharov and Shabat rigorously
established that the NLSE admits one-dimensional solitary-wave (soliton)
solutions that preserve their temporal/spectral profiles during propagation
and after mutual collisions \cite{shabat1972exact}. In this pioneer study,
the consideration of NLSE was inspired by previous work on plasma physics
and hydrodynamics. Soon afterwards, Hasegawa and Tappert recognized that
optical fibers, through the interplay of the group-velocity dispersion (GVD)
and Kerr nonlinearity, provide a natural medium to realize both bright \cite%
{hasegawa1973transmission_a} and dark \cite{hasegawa1973transmission_b}
solitons, thus predicting that the solitons, in the form of intense
picosecond pulses, could propagate essentially undistorted over long
distances. This prediction was confirmed in 1980 by the pioneering
experiments performed by Mollenauer, Stolen, and Gordon in single-mode
silica fibers \cite{mollenauer1980experimental}.

\begin{figure}[t]
\centering
\includegraphics[width=12cm]{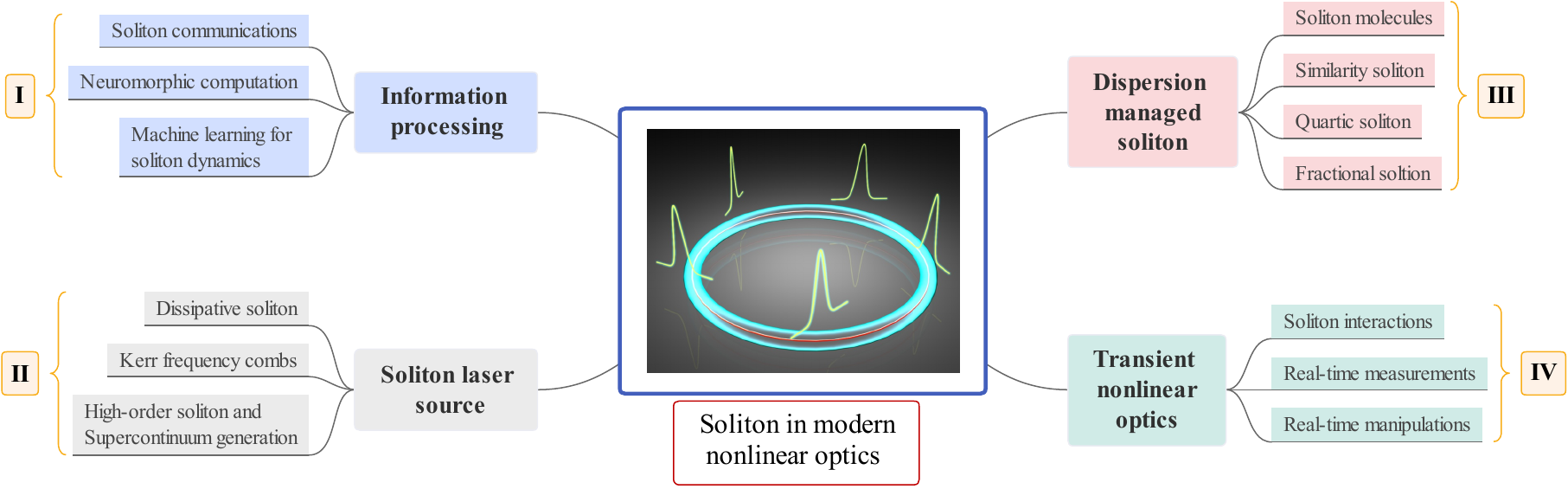}
\caption{Optical solitons across fundamental science and applications.
\textbf{I}: Information processing. \textbf{II}: Soliton laser sources.
\textbf{III}: Dispersion-managed solitons. \textbf{IV}: Transient nonlinear
optics and real-time diagnostics.}
\label{F1}
\end{figure}

These milestone results had established solitons as a central paradigm in
nonlinear fiber optics and launched extensive research into their physics
and applications. The ability of the solitons to maintain their shape and
coherence over long propagation spans made them compelling candidates for
overcoming dispersive penalties in optical communications (see \textbf{%
Branch I} in Fig.~\ref{F1}). From the late 1980s onward, the reported
transmission over hundreds and then thousands of kilometers demonstrated the
feasibility of soliton-based telecommunication links \cite%
{mamyshev1991generation,hasegawa2000historical,hasegawa2002soliton,mollenauer2006solitons}%
. With the help of the \Red{dispersion management (DM)} \cite{turitsyn2012dispersion} and coherent
detection, which later became the mainstream for the design of high-capacity
systems \cite{sano200240,weber2008ultra}, the soliton concept helped to
establish the broad principle that nonlinearity can be \emph{harnessed} -
rather than suppressed - for the long-distance data transmission in optical
fibers. Beyond the realm of telecommunications, solitons underpin a wide
range of wave phenomena in optical fibers, getting increasingly integrated
into optical computation systems (e.g., neuromorphic photonics) \cite%
{hary2025principles,genty2025computing} and data-driven prediction
frameworks \cite%
{marcucci2020theory,guo2021integrated,fischer2023neuromorphic,genty2021machine,ma2023artificial}%
. For a historical overview of the topic, see \cite%
{dudley2023fifty,blanco2023bright}; for a broad exposition of the
background, consult the book \emph{Nonlinear Fiber Optics} by Agrawal \cite%
{agrawal2000nonlinear}.

In ultrafast laser science, solitons underpin the operation of passively
mode-locked fiber lasers (\textbf{Branch II} in Fig.~\ref{F1}). The
interplay of dispersion, Kerr nonlinearity, gain, and spectral filtering in
a resonator naturally supports the formation of solitons, providing compact
and robust sources of femtosecond optical pulses \cite%
{haus2002mode,keller2003recent,wise2008high,grelu2012dissipative}. The
concept of \emph{dissipative} solitons has greatly expanded the accessible
pulse parameter space (energy, duration), facilitating the creation of
high-energy pulses in picosecond and femtosecond ranges, which are used to
seed highly nonlinear processes. A prominent example is the SC generation,
where fission of the higher-order-soliton and dispersive-wave emission
produce octave-spanning spectra \cite%
{dudley2006supercontinuum,alfano2006supercontinuum}, with applications to
spectroscopy \cite{borondics2018supercontinuum}, microscopy \cite%
{kaminski2008supercontinuum,gundougdu2023self}, and optical coherent
tomography \cite{tu2013coherent,ji2021millimeter}. Further, temporal cavity
solitons in Kerr microresonators underlie the operation of chip-scale
frequency combs for precision metrology, microwave photonics, and emerging
quantum technologies \cite%
{herr2014temporal,trocha2018ultrafast,herr2016dissipative}.

Beyond the classical dissipative-soliton paradigm, experimental and
theoretical developments have revealed a rich variety of \emph{%
dispersion-managed} soliton states in fibers and \Red{waveguides} \cite{turitsyn2012dispersion} (\textbf{Branch III}). \emph{Soliton molecules} (SMs) --
bound states of two or several temporal optical solitons -- offer new
degrees of freedom (the temporal separation and relative phase between the
bound solitons) for dense information encoding and exploring nonlinear
inter-soliton coupling \cite%
{malomed1992bound,tang2002bound,stratmann2005experimental,malomed2005bound,
song2019recent,liu2022demand,liu2023phase,cui2026bound}, Higher-order
GVD enables to establish \emph{quartic} solitons in regimes dominated by the
fourth-order dispersion (FOD), which feature a distinctive energy- duration
scaling \cite{blanco2016pure,runge2020pure}. In parallel, \emph{fractional}
solitons , which are represented by solutions of fractional-order extensions
of the NLSE (FNLSE), have been realized in a mode-locked fiber laser system
\cite{liu2025experimental,hoang2025nonlinear}. The fractional solitons open
a new frontier in nonlocal wave dynamics \cite%
{laskin2002fractional,fujioka2010fractional,malomed2021optical}. These
advancements expand the understanding of nonlinear wave physics and suggest
applications to multidimensional encoding, advanced pulse synthesis, and
analog simulation of exotic quantum models.

A defining aspect of fiber-soliton physics is the ultrafast, \emph{transient}
nature of nonlinear dynamics (\textbf{Branch IV }in Fig.~\ref{F1}). Solitons
can interact, collide, pulsate, or undergo explosions on femtosecond and
picosecond scales. Real-time diagnostics have been crucial to observe such
processes directly. The dispersive Fourier transform (DFT) maps single-shot
spectra into the time domain, enabling round-trip-resolved tracking of
events such as interactions, explosions, and rogue-wave formation \cite%
{solli2007optical,goda2013dispersive,dudley2014instabilities,herink2017real,mahjoubfar2017time}%
. Complementarily, the time-lens technique uses quadratic temporal phase to
realize temporal imaging with the sub-picosecond resolution \cite%
{kolner2002space,salem2013application,zhang2025ultra}. Together, these
methods link the theory and experiment and reveal the diverse character of
the evolution of temporal solitons in optics.

\paragraph{The scope of this review.}

We survey soliton generation and control in modern mode-locked fiber-laser
systems. We first address dissipative solitons (DSs) with an emphasis on the
control of the pulse duration and energy (in particular, for driving
strongly nonlinear processes such as SC). We then review advanced single-pulse soliton states enabled by dispersion and cavity engineering, including dispersion-managed solitons, similaritons, vector solitons, pure-quartic solitons, and fractional solitons, followed by multi-pulse bound states, including stationary and breathing soliton molecules. Real-time measurements and manipulation strategies for DM
solitons are considered too. Finally, we review recent advancements in
\textquotedblleft smart" soliton lasers, leveraging the intelligent
search/feedback strategies. Throughout the chapter we aim to highlight
connections between theoretical models, experimental platforms, and emerging
applications.

\Blue{To maintain a consistent terminology throughout this review, we classify the
different soliton states primarily according to their governing propagation
models, dispersion regimes, and dominant physical balances. In this framework,
conventional solitons arise predominantly from the balance between anomalous
group-velocity dispersion and Kerr nonlinearity, whereas dissipative solitons
are sustained by a broader balance involving dispersion, nonlinearity, gain,
loss, and spectral filtering. Dispersion-managed states are distinguished by
their periodic evolution under alternating or engineered dispersion, and may
include self-similar, higher-order-dispersion, and fractional-dispersion
soliton regimes. Multi-pulse bound states, referred to here consistently as
soliton molecules, are characterized additionally by stable or dynamically
evolving temporal separations and relative phases resulting from
inter-soliton interactions. Accordingly, the various soliton families
discussed below are compared, whenever applicable, in terms of their governing
equations and characteristic parameters, formation conditions, propagation
and interaction dynamics, methods of manipulation and control, and potential
applications. This unified framework is intended to clarify both the
distinctions and the physical connections among the diverse nonlinear wave
states encountered in modern fiber-laser systems.}

\section{Soliton-State Family Generation in Fiber-Laser Systems}

\label{sec:families}

\begin{figure}[t]
\centering
\includegraphics[width=12cm]{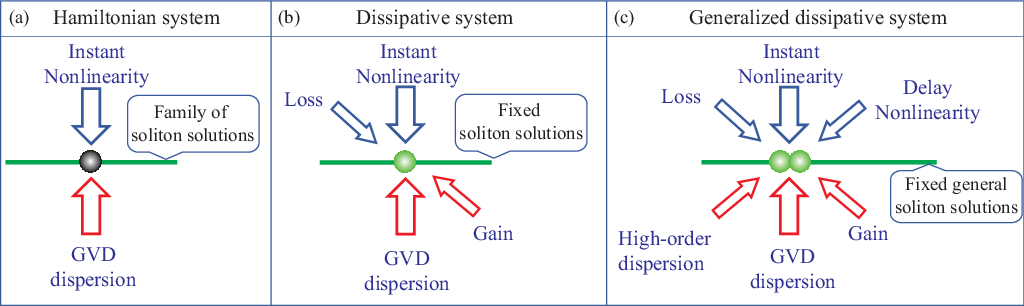}
\caption{Soliton-state families in nonlinear optical systems. (a) The
Hamiltonian system (it realizes the conservative balance of GVD and
instantaneous Kerr nonlinearity). (b) The dissipative system (with the
simultaneous balance of GVD, nonlinearity, gain, and loss). (c) The
generalized dissipative system including higher-order GVD and delayed
nonlinear response, together with the spectral filtering and distributed
gain/loss. The plot is redrawn and borrowed from work \cite{grelu2012dissipative}.}
\label{F2}
\end{figure}

Soliton formation in nonlinear optics -- particularly, in optical fibers --
can be broadly categorized into three classes of systems according to the
underlying balance of physical effects. In \emph{Hamiltonian (conservative)
systems} [Fig.~\ref{F2}(a)], the dynamics are governed by the nonlinear Schr%
\"{o}dinger equation (NLSE), in which the GVD and \emph{instantaneous} Kerr
nonlinearity may exactly compensate each other. The balance supports a
continuous family of soliton solutions parameterized by the pulse's energy
or duration, as established in the pioneering works of Zakharov and Shabat
and subsequently realized theoretically in optical fibers by Hasegawa and
Tappert \cite%
{shabat1972exact,hasegawa1973transmission_a,hasegawa1973transmission_b,mollenauer1980experimental,agrawal2000nonlinear}%
. If the NLSE includes solely the second-order GVD and cubic self-phase
modulation (SPM) term, the equation is integrable and, accordingly, the
soliton dynamics obey exact analytical solutions \cite{shabat1972exact}.

In \emph{dissipative systems} [see Fig.~\ref{F2}(b), which is drawn
following the conceptual scheme of Ref.~\cite{grelu2012dissipative}], which
are typical for passively mode-locked fiber lasers and driven Kerr cavities
and microresonators, the soliton existence results from the \emph{%
simultaneous} balance of the Kerr nonlinearity, GVD, gain, and loss. Unlike
the Hamiltonian case, DSs do not form families, but represent isolated \emph{%
attractor} \emph{solutions}, whose parameters are uniquely determined by the
cavity (gain, loss, spectral filtering) \cite%
{chang2008dissipative,grelu2012dissipative,herr2016dissipative}.

In practice, most experimental platforms fall into a more complex regime
that we term the \emph{generalized dissipative system} [Fig.~\ref{F2}(c)].
Here, the basic dissipative balance is perturbed by higher-order GVD,
dispersive losses (spectral filtering), delayed nonlinear response (e.g.,
Raman self-frequency shift), Brillouin/Raman scattering, self-steepening,
and distributed gain/loss in different segments of the cavity. In Fig.~\ref%
{F2}(c) we explicitly add the higher-order GVD and delayed nonlinearity to
emphasize the presence of these factors \cite%
{agrawal2000nonlinear,sun2024multimode}. They essentially affect advanced
soliton states, including \emph{SMs} \cite{malomed2005bound,herink2017real},
\emph{quartic solitons} in the FOD systems \cite{blanco2016pure}, \emph{%
fractional-order solitons} in the nonlocal (fractional-GVD) setting \cite%
{laskin2002fractional,malomed2024basic,liu2025experimental}, and \emph{%
spatiotemporal} solitons in multimode or transversely structured fibers \cite%
{silberberg1990collapse,malomed2005spatiotemporal,sun2024multimode,dudley2006supercontinuum}%
. The generalized dissipative framework thus provides a comprehensive
setting for the realization of the fiber-based nonlinear dynamics, linking
the idealized Hamiltonian soliton picture to the rich variety of DS families
observed experimentally in modern mode-locked fiber-laser systems.

\subsection{DS families: engineering pulse duration}

\label{subsec:DS_engineering}

The mode-locked fiber laser (MLFL) has become a paradigmatic platform for
the DS generation, with the interplay of the GVD, Kerr nonlinearity, gain,
and loss being naturally established in a resonant fiber cavity. A typical configuration is shown in Fig.~\ref{F3}(a):
an erbium-doped fiber (EDF) provides optical gain; a wavelength-division
multiplexer (WDM) couples the pump; a Saturable Absorber (SA) or \Red{Nonlinear Polarization Rotation (NPR)} section
enables passive mode locking; an
output coupler (OC) extracts a fraction of the intracavity power; and a
length of single-mode fiber (SMF) sets the net cavity dispersion.
Intracavity components, such as spectral filters and isolators (ISO),
further tailor the bandwidth and unidirectionality of the pulse evolution.
Within this architecture, stable \emph{attractor} \emph{pulses} readily
produce DSs, in contrast to continuous families of conservative solitons in
Hamiltonian models \cite{grelu2012dissipative,agrawal2000nonlinear}. The
pulse duration, as the most important DS parameter, is strong affected by
the MLFL\ design. This section focuses, below,on three main types of MLFLs
to engineer soliton pulses with different durations.

\subsubsection{MLFLs (mode-locked fiber lasers) and fundamental DSs
(dissipative solitons)}

\label{subsubsec:MLFL_fundamental_DS}

The pulse dynamics in MLFLs are accurately modeled by the complex
Ginzburg--Landau equation, which augments the NLSE with distributed
gain/loss and spectral filtering \cite%
{soto1996stability,akhmediev1997multisoliton,grelu2012dissipative}. For the
simulations and design, it is often advantageous to model each subsections
of the cavity separately and concatenate their actions per round trip. In
the nonlinear gain-carrying fiber segments, the slowly varying envelope $%
\Psi (t,z)$ is commonly described by the NLSE,
\begin{equation}
i\,\frac{\partial \Psi }{\partial z}=\frac{\beta _{2}}{2}\,\frac{\partial
^{2}\Psi }{\partial t^{2}}+i\,\frac{g(z)}{2}\,\Psi -\gamma \,|\Psi |^{2}\Psi
,  \label{E-NLSE}
\end{equation}%
where $z$ is the propagation distance, $t=\mathrm{t}-\beta _{1}z$ is the
retarded time ($\mathrm{t}$ is the original time, and $\beta _{1}$ is the
group velocity of the carrier wave), $\beta _{2}$ is the GVD coefficient, $%
g(z)$ represents the distributed linear gain ($>0$) or loss ($<0$), and $%
\gamma >0$ is the Kerr coefficient. If necessary, higher-order GVD terms can
be added to Eq. \eqref{E-NLSE} \cite{karpman1993radiation,vithya2018combined}%
.

For the soliton operation in the telecom band, the gain medium is typically
provided by Er doped fiber. An appropriate phenomenological model for the Er doped gain
saturation is
\begin{equation}
g(z)=g_{0}\,\exp \!\left[ -\frac{E(z)}{E_{\mathrm{sat}}}\right] ,
\label{E-gain}
\end{equation}%
where $g_{0}$ is the linear (small-signal) gain, $E_{\mathrm{sat}}$ the
saturation energy, and
\begin{equation}
E(z)=\int_{-\infty }^{+\infty }|\Psi (z,t)|^{2}\,dt  \label{E-energy}
\end{equation}%
is the total energy of the pulse.

Stable mode locking in MLFL is enforced by an intracavity SA, which may be
implemented by SESAMs (semiconductor saturable absorber mirrors) \cite%
{keller1996semiconductor,kopf1996all}, nonlinear-loop mirrors/nonlinear
amplifying loop mirror (NOLM/NALM) \cite{fermann1990nonlinear,kuse2016all},
or by artificial SAs based on nanomaterials, such as carbon nanotubes and
graphene \cite%
{set2004laser,hasan2009nanotube,martinez2010optical,huang2020dynamics,lau2022comparison}%
.  A commonly used intensity-transfer model for a fast SA is
\begin{equation}
T_{\mathrm{SA}}(I)\;=\;\sqrt{\,1-\left( \alpha _{\mathrm{ns}}+\frac{\alpha
_{0}}{1+I/I_{\mathrm{sat}}}\right) \,}\,,  \label{E-SA}
\end{equation}%
where $I=|\Psi |^{2}/A_{\mathrm{eff}}$ is the instantaneous intensity, $A_{%
\mathrm{eff}}$ the effective mode area, $\alpha _{\mathrm{ns}}$ the
non-saturable loss, $\alpha _{0}$ the saturable loss, and $I_{\mathrm{sat}}$
the saturation intensity. Slow SAs can be modeled more accurately with a
dynamical population equation including the \Red{recovery time} \cite{haus1975theory,paschotta2001passive}.

\begin{figure}[t]
\centering
\includegraphics[width=12cm]{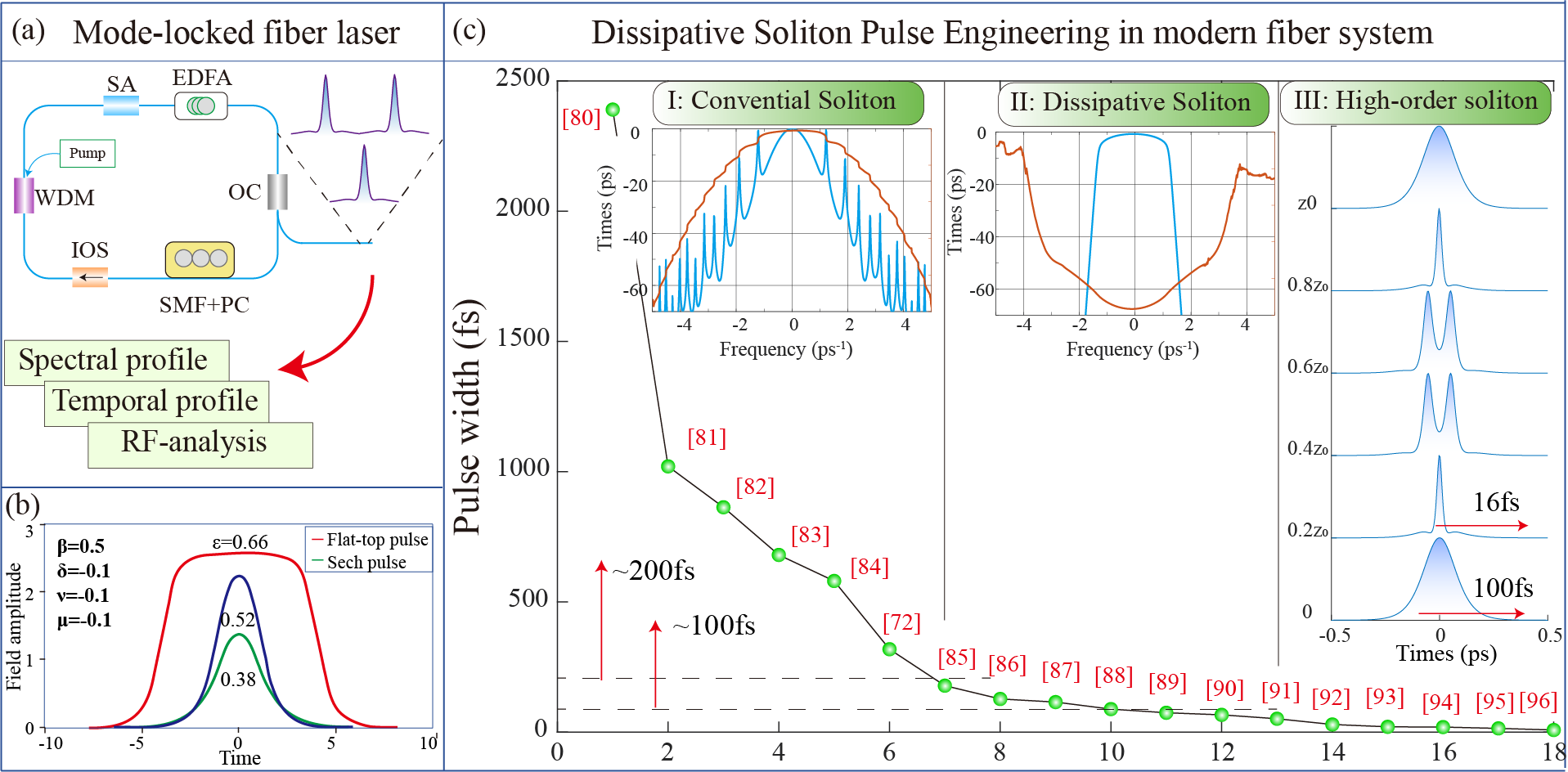}
\caption{(a) A typical MLFL setup (EDF/WDM/SA or NPR/OC/SMF with optional
filter and ISO). (b) The temporal intensity of the DS (redrawn from Ref.~\protect\cite%
{grelu2012dissipative}). (c) An example of the spectrum and phase of steady states from
simulations of Eqs.~\eqref{E-NLSE}--\eqref{E-SA}: I) a conventional soliton
(net anomalous GDD), II) a DS (net normal GDD) and III) a high-order soliton.
Here, green dots connected by the curve refer to pulse durations
(around 1550\,nm) spanning $\sim$2\,ps to $\sim$8\,fs \protect\cite%
{wang2008wideband,song2008single,martinez2011multi,huang20192,
nicholson2007optically,set2004laser,rozhin2006sub,lazarev2016stable,
shohda2008passively,sotor2015sub,popa201274,yu201366,soltanian2025all,
purdie2015few,kibler2006all,liu2025engineering,kieu2010generation,sell20098}%
, which are also marked in (c). }
\label{F3}
\end{figure}

Depending on the balance between the GVD, nonlinearity, gain saturation,
spectral filtering, and SA dynamics, the resulting steady-state pulses range
from near-sech conventional solitons to flat-top and strongly chirped DSs,
see Fig.~\ref{F3}(b). Recent reviews provide comprehensive taxonomies of
these regimes and transitions between them \cite%
{grelu2012dissipative,song2019recent,huang2020dynamics,grelu2024solitary}.

\paragraph{Engineering and benchmarking the pulse duration.}

In terms of the fiber-laser operation, the pulses's temporal width
(duration) $T$ is the central figure of merit, as it sets
dispersion/nonlinearity lengths and compression ratios. In MLFLs, durations can be tailored
from picoseconds to femtosecond---and even few-cycle---regime by the tuning
of: (i) the net cavity dispersion (normal vs.\ anomalous); (ii) the
spectral-filter center/bandwidth; (iii) the gain map and pump level; and
(iv) the SA/NPR parameters.

The green dots in Fig.~\ref{F3}(c) indicate values of the pulse duration as
extracted from the references listed in the caption, for the central
wavelength around $1550$\thinspace nm \cite%
{wang2008wideband,song2008single,martinez2011multi,huang20192,
nicholson2007optically,set2004laser,rozhin2006sub,lazarev2016stable,
shohda2008passively,sotor2015sub,popa201274,yu201366,soltanian2025all,
purdie2015few,kibler2006all,liu2025engineering,kieu2010generation,sell20098}%
. Two boundary values of the pulse width, \textit{viz.}, $T =100$ fs and $%
200 $ fs, are marked in Fig.~\ref{F3}(c), thus defining three regions of
practical significance. For $T \geq 200~\mathrm{fs}$ (region \textbf{I}),
MLFLs operating with the \emph{anomalous} net GDD maintain conventional
solitons. In the intermediate region \textbf{II}, i.e., $100$ $\mathrm{fs}<T
<~200~\mathrm{fs}$, DSs, supported by external pulse compression, typically
occur. At $T <100$ $\mathrm{fs}$ (region \textbf{III}, down to the few-cycle
regime), higher-order soliton dynamics, nonlinear spectral broadening with
compression, and supercontinuum generation should be taken into regard.

\paragraph{Region I: conventional solitons (net anomalous dispersion).}

With negative net group-delay dispersion (GDD), the cavity supports pulses
which are similar to the sech-shaped classical NLSE\ (Hamiltonian) solitons.
Their spectra typically exhibit \emph{Kelly sidebands} arising from resonant
generation of dispersive waves by the soliton in the periodic gain/loss
landscape \cite{kelly1992characteristic}. A representative spectrum is
inserted in region I of Fig.~\ref{F3}(c) (the top left panel), obtained from
simulations based on Eqs.~\eqref{E-NLSE}--\eqref{E-SA} with net GDD $\approx
-0.15\,\mathrm{ps^{2}}$. Experimentally, this regime routinely delivers
pulses with $T$ from $\sim \!2$\thinspace ps to $\sim \!200$\thinspace fs
pulses, with the intracavity duration limited by soliton-energy constraints.

\paragraph{Region II: DSs (net normal dispersion).}

For small positive net GDD (e.g., $+0.01\,\mathrm{ps^{2}}$ for the spectrum
in region II in the top medium panel of Fig.~\ref{F3}(c)), the steady state
is a strongly chirped DS with a characteristically rectangular spectrum and
broad bandwidth. The residual positive spectral phase enables efficient
\emph{external} compression to the near-transform-limited pulse width \cite%
{lazarev2016stable}. When the cavity implements an all-normal-dispersion map
with appropriate gain and filtering, self-similar (parabolic) pulses and
similaritons may emerge, offering high-energy, highly compressible outputs
\cite{fermann2000self,ilday2004self,dudley2007self} (%
\textbf{also see the section of DM soliton for
similariton}).

\paragraph{Region III: High-order soliton.}

Pushing $T$ below $\sim \!100$\thinspace fs demands wider bandwidth. In
practice, the extension in the bandwidth is usually implemented by means of
out-of-cavity SPM effect in the fiber system. In this manner, an external EDFA system
is usually employed to amplify the pulse energy and thus to form a
high-order soliton. In this regime, the pulse may be compressed into a
few-cycle regime.

\subsubsection{High-order soliton and SC (supercontinuum)}

To further reduce the pulse width $T$ into Region III, one needs to produce
the soliton with a wider spectrum. High-order solitons provide a useful
resource to achieve this purpose, They are characterized by periodic
temporal and spectral evolution, which is determined by the soliton order $N$
\cite{agrawal2000nonlinear}. For $N>1$ (defined in the bottom line of Eq. %
\ref{length}), the pulse undergoes periodic compression and broadening in
the course of the propagation, enabling the generation of ultrashort pulses,
with duration $T$ far below the input width. Near the compression point,
few-cycle pulses may be produced, limited only by the fiber's dispersion map
and the available nonlinearity \cite{dudley2006supercontinuum}. The panel in
the right top side of Fig. \ref{F3}(c) displays simulations for the pulse
evolution of the high-order soliton ($N=3$), in which the incident-pulse's
width is $T=100$ fs, is compressed to the pulse with the minimum width $16$
fs achieved after the propagation distance $0.2z_{0}$. Here, the soliton's
number, $N$, and soliton period, $z_{0}$, are determined by the dispersion
and nonlinearity length \cite{agrawal2000nonlinear}, \textit{viz}.,

\begin{equation}
\begin{split}
L_{\mathrm{d}}& =\frac{T_{0}^{2}}{|\beta _{2}|},\quad \text{(Dispersion
Length)} \\
L_{\mathrm{n}}& =\frac{1}{\gamma P_{\mathrm{k}}},\quad \text{(Nonlinearity
Length)} \\
z_{0}& =\frac{\pi L_{\mathrm{d}}}{2},\quad \text{(Soliton Period)} \\
N^{2}& =\frac{L_{\mathrm{d}}}{L_{\mathrm{n}}},\quad \text{(Soliton Number)}
\end{split}
\label{length}
\end{equation}%
where $P_{\mathrm{k}}$ is the peak power of the initial pulse, and $T_{0}$
is its duration. The input field is taken as $\Psi _{0}(t)=\sqrt{P_{\mathrm{k%
}}}\,\mathrm{sech}(T/T_{0})$. By carefully adjusting the dispersion and
nonlinearity length of the incident pulse, one can observe the dynamics of
the $N$-th order solitons in the framework of the NLSE \cite%
{mollenauer1980experimental,hodel1987decay,wai2003ultrashort,liu2025engineering}%
.

In realistic systems, higher-order solitons are perturbed by higher-order
GVD, the Raman scattering and self-steepening, leading to the soliton
fission and emission of dispersive waves. This process is the physical basis
for the SC generation, where a narrowband input pulse is converted into an
ultrabroad optical spectrum spanning hundreds of nanometers. In this case,
the model is provided by the generalized NLSE \cite{dudley2006supercontinuum}%
,
\begin{equation}
\begin{split}
& \qty[\pdv{z}-i\sum_{k\geq2}\frac{i^k\beta _k}{k!}\pdv[k]{t}+\frac\alpha2]%
\Psi (z,t)= \\
& \gamma \qty(i-\tau _{\mathrm{shock}}\pdv{t})\qty[i\Gamma_R(z,t)+\Psi(z,t)%
\int_{-\infty}^{+\infty}dt'\;R(t')\abs{\Psi(z,t-t')}^2].
\end{split}
\label{E-SC}
\end{equation}%
The left-hand side of this equation models linear propagation effects, while
the right-hand side accounts for nonlinear effects, with the Kerr
coefficient $\gamma $ representing the SPM term, and $R(t)$ being the Raman
response function. Term $\Gamma _{R}(z,t)$ represents the spontaneous Raman
noise. Additional nonlinear phenomena, such as the self-steepening and
optical-shock formation, are typically characterized by a timescale $\tau _{%
\mathrm{shock}}=1/\omega _{0}$.

The SC generation in photonic-crystal fibers and highly nonlinear fibers has
been widely exploited for producing femtosecond and even few-cycle pulses,
with the help of an appropriate dispersion-compensation scheme \cite%
{dudley2006supercontinuum,skryabin2010colloquium}. Typically, the pulse
width $T \simeq 8$ ps was achieved by passing an MLFL-generated soliton
through a highly nonlinear fiber \cite{sell20098}. In ultrafast photonics,
the SC generation and precise control of the respective spectro-temporal
features provides breakthroughs in the development of ultrashort laser
sources, including self-referenced \cite%
{Jones2000Sci,brida2014ultrabroadband,Sinclair2015comb} and offset-free \cite%
{Fehrenbacher2015Optica} frequency combs, the generation of few- and
single-cycle pulses \cite%
{Krauss2010NatPhot,shumakova2016multi,steinleitner2022single}, and the
high-harmonic generation by means of attosecond pulses. \cite%
{Lewenstein1994PRA,Corkum2007NatPhys,Krausz2024RMP,peng2019attosecond,li2020attosecond}%
.

\section{Dispersion- and Cavity-Engineered Single-Pulse Soliton States}

\begin{figure}[tbp]
\centering
\includegraphics[width=12cm]{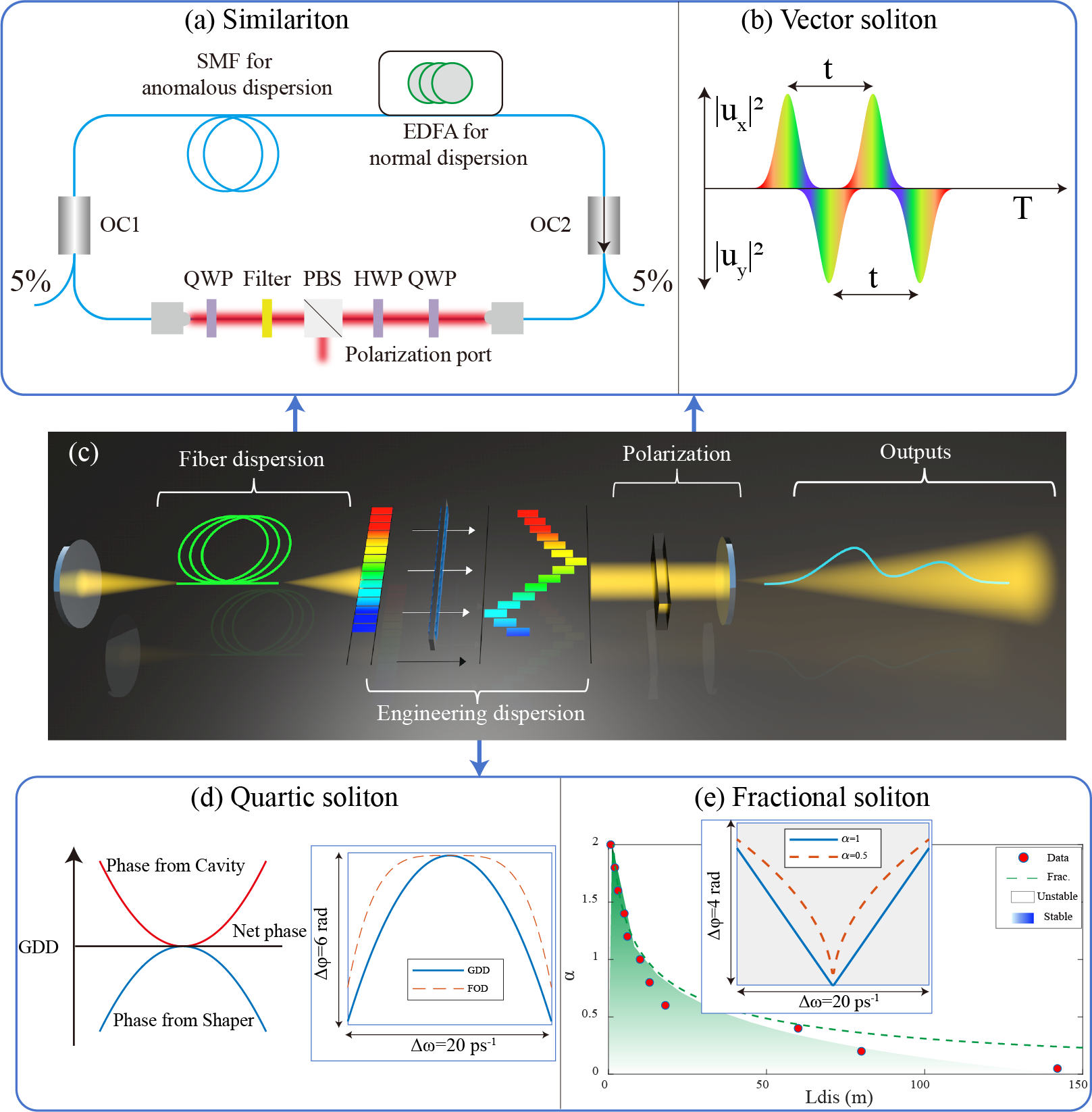}
\caption{Overview of DM soliton states in MLFLs. (a) A \emph{similariton},
represented by the measured/simulated intra-cavity evolution in the Er-doped
laser, which features strong spectral breathing as the pulse circulates in
the cavity. The figure is borrowed from Ref. \protect\cite{oktem2010soliton}%
. (b) A \emph{vector soliton}: SMs (soliton molecules) in
birefringence-managed fiber lasers, as per Ref. \protect\cite%
{he2023dislocated}. (c) The \emph{dispersion management }(DM) concept: the
net cavity GVD induced by fiber segments and space-domain-engineering
dispersion (a grating/prism or programmable pulse shaper), with the
polarization control. (d) A \emph{pure quartic soliton (}PQS): the
phase from the cavity and pulse shaper system at the left panel; with the right panel showing the
effective FOD (fourth-order dispersion) and GDD. These plots are borrowed from Ref. \cite{blanco2016pure}. (e) A \emph{fractional
soliton}: the stability map in the plane of the L\'{e}vy index $\protect%
\alpha $ and fractional-dispersion distance $L_{\mathrm{dis}}$ (green: the
actual stability region, while the dashed curve indicates the estimated
stability boundary); this panel is redrawn from \cite{liu2025experimental}. Inset: an example of fractional-GVD phases for $\protect%
\alpha =1$ and $0.5$. }
\label{F5}
\end{figure}

The interplay of GVD with the Kerr nonlinearity governs the pulse dynamics
in MLFLs. Owing to the distributed actions of the GVD and nonlinearity in
different cavity segments [see Fig.~\ref{F5}(c)], the intracavity pulse
exhibits a characteristic spectrotemporal evolution in each round trip. In
practice, dispersion is easier to engineer than the intrinsic nonlinearity;
consequently, appropriate adjustment of the cavity GVD enables formation of
DM solitons. As sketched in Fig.~\ref{F5}(c), the net GVD arises from fiber
segments and from the \emph{space-domain} dispersion introduced by a pulse
shaper or other spatiotemporal elements.

Below, following the GVD and polarization engineering in cavities, we
introduce four representative single-pulse states: a \emph{self-similar
soliton} (similariton), a \emph{vector} \emph{soliton}, a \emph{quartic}
\emph{soliton}, dominated by FOD, and a \emph{fractional} \emph{soliton, }%
governed by effective fractional-order GVD.

\subsection{Similaritons}

\label{subsubsec:similariton}

As discussed in Sec.~\ref{subsubsec:MLFL_fundamental_DS}, the soliton
species in MLFLs can be identified as: (i) \emph{conventional solitons}
(corresponding to \emph{negative} net GDD), (ii) \emph{dissipative solitons}
(corresponding to \emph{positive} net GDD), and (iii) \emph{self-similar
solitons} (similaritons), when the net GDD nearly vanishes.

Before turning to cavity similaritons, it is relevant to recall the
self-similar pulse evolution in fiber amplifiers with the \emph{normal GVD}.
High-energy pulse propagation under the action of normal GVD can avoid wave
breaking, provided that the pulse maintains a parabolic temporal intensity
profile and a quadratic temporal phase (i.e., a linear chirp), as first
predicted by Anderson \cite{anderson1993wave}. Subsequent numerical and
experimental studies have shown that arbitrarily shaped input pulses evolve
toward ones with such parabolic, linearly-chirped shapes, which represent an
asymptotic solution of NLSE with the gain term \cite%
{tamura1996pulse,fermann2000self,kruglov2002self,dudley2007self}. In the
pioneering work by Fermann \cite{fermann2000self}, \Red{the self-similar solution
to the gain-augmented NLSE was obtained as
\begin{equation}
\begin{aligned} \Psi(z,T) &= \Psi_0(z)\sqrt{1 -
[T/T_0(z)]^2}\,\exp\!\big(i\phi(z,T)\big), \\ \phi(z,T) &= \phi_0 +
\frac{3\gamma}{2g}\,A_0^2(z) - \frac{g}{6\beta_2}\,T^2, \end{aligned}
\label{E-self-sim}
\end{equation}}%
where, in the asymptotic limit of $z\rightarrow \infty $, the effective
width $T_{0}(z)$ and amplitude $\Psi _{0}(z)$ take the form of
\begin{equation}
\begin{aligned} T_0(z) &=
3\,g^{-2/3}\,(cE_{\mathrm{IN}})^{1/3}\,\exp\!\left(\frac{gz}{3}\right),\\
\Psi_0(z) &=
\tfrac{1}{2}\,(gE_{\mathrm{IN}})^{1/3}\,c^{-1/6}\,\exp\!\left(\frac{gz}{3}%
\right), \end{aligned}  \label{E-self-sim2}
\end{equation}%
with $c=\gamma \beta _{2}/2$, input energy $E_{\mathrm{IN}}$, and gain
parameter $g$. The respective instantaneous frequency,
\begin{equation}
-\frac{\partial \phi (z,T)}{\partial T}\;=\;\frac{g}{3\beta _{2}}\,T,
\label{E-chirp}
\end{equation}%
is a linear function of $T$, while $|\Psi (z,T)|^{2}$ is parabolic.
The linear chirp enables efficient compression, using standard dispersive
compressors (e.g., gratings). In that pioneering work \cite{fermann2000self}%
, launching a 200~fs (FWHM) pulse at $1.06\,\mathrm{\mu }$m into a Yb-doped
fiber amplifier and compressing the pulse externally yielded its temporal
width $T\approx \!68$~fs (with the theory indicating the possibility
reducing the width up to $T\simeq \!30$~fs).

These properties have direct implications for lasers. Stable self-similar
pulses can exist at energies much higher than allowed by soliton-like
shaping, motivating cavity designs that exploit normal GVD, as demonstrated
by Ilday \emph{et al.} \cite{ilday2004self}. In their Yb:fiber laser, normal
GVD was provided by SMF, while a dispersive delay line supplied anomalous
GVD. The gain bandwidth acted as an effective spectral filter, creating a
self-consistent resonator solution. Chirped-pulse energies of $\sim \!10$~nJ
were achieved (about three times that of a DM-soliton laser at the net
dispersion $0.003\,\mathrm{ps^{2}}$). Similar regimes have since been
realized in Yb-doped lasers \cite%
{nielsen2005self,ortacc2006self,ma2019high,teugin2019spatiotemporal} and in
Er-doped systems \cite%
{ruehl20050,oktem2010soliton,peng2012experimental,wang2016generation,peng2016gain}%
.

Figure~\ref{F5}(a) illustrates \Red{the setup for the pulse dynamics} in the Er-doped fiber
laser implementing the similariton operation \cite{oktem2010soliton}. The
EDF provides normal GVD (supporting the self-similar amplification), while
the SMF furnishes anomalous GVD. A saturable absorber and a spectral filter
between them reduce the temporal and spectral width, helping to create an
effective attractor for similaritons. Experimentally, spectral breathing ratios up to $\sim \!9$
were observed for an $8$~nm filter bandwidth with net dispersion $\approx
0.0136\,\mathrm{ps^{2}}$.

We emphasize the distinction between \emph{DSs} and \emph{similaritons} in
the normal-dispersion regime. As said above, DSs are steady-state attractors
of the driven cavity, whereas similaritons are evolutionary states
approaching a parabolic asymptote. Both support high energies (in
particular, the DSs energy may achieve $\simeq \!30$~nJ \cite%
{kieu2009sub,jeong2014all,tang2015high}), but they are maintained by
different balances, \textit{viz}., the static and self-scaling ones,
respectively. Hallmark features of similaritons are (i) strong spectral
breathing (typically, with a respective factor $\sim \!10$) \cite%
{oktem2010soliton} and (ii) excellent pulse quality with parabolic temporal
profiles and strong linear chirp, permitting near-transform-limited
compression with minimal wings. For comprehensive overview of the topic, see
Refs. \cite{dudley2007self,chong2015ultrafast,woodward2018dispersion}.

\subsection{Vector soliton}

\label{subsubsec:vector soliton}

From the theoretical viewpoint, \emph{vector solitons} are multi-component
localized wave packets predicted as solutions of systems of coupled
nonlinear evolution equations, in which each component is affected not only
by SPM but also by inter-component interaction, most notably cross-phase
modulation (XPM), and, in some cases, coherent four-wave-mixing terms \cite%
{menyuk2003nonlinear,agrawal2000nonlinear}. In fiber systems, the most
common realization is the polarization vector soliton, formed by two
orthogonally polarized field components propagating in a weakly birefringent
medium or cavity. Unlike scalar solitons, whose existence is determined
primarily by the balance between the GVD and Kerr nonlinearity, vector
solitons arise from the joint balance involving the GVD, birefringence,
nonlinear coupling, and, in laser cavities, dissipative effects such as the
gain, loss, saturable absorption, and spectral filtering \cite%
{akhmediev1999partially,grelu2012dissipative}.

The basic theoretical description of polarization vector solitons is
provided by the system of coupled nonlinear Schr\"{o}dinger equations for
the two orthogonally polarized envelopes $u(z,t)$ and $v(z,t)$:
\begin{equation}
\begin{aligned} i\frac{\partial u}{\partial z}
+\frac{\beta_2}{2}\frac{\partial^2 u}{\partial t^2} +\gamma
\left(|u|^2+\kappa |v|^2\right)u +cu^{*}v^2e^{-2i\Delta \beta z}&= 0,\\
i\frac{\partial v}{\partial z} +\frac{\beta_2}{2}\frac{\partial^2
v}{\partial t^2} +\gamma \left(|v|^2+\kappa |u|^2\right)v
++cv^{*}u^2e^{2i\Delta \beta z} &= 0, \end{aligned}  \label{E-vector}
\end{equation}%
where $\kappa $ and $c$ are the relative strengths of XPM and four-wave
mixing, respectively. For the linearly birefringent fiber, $\kappa =2/3$ and
$c=1/3$ \cite{agrawal2000nonlinear}. In passively mode-locked fiber lasers,
this conservative model is further extended to the system of coupled complex
Ginzburg--Landau equations, which account for\ the gain saturation, cavity
loss, spectral filtering, and saturable absorption \cite%
{song2019recent,grelu2012dissipative}. In this sense, vector solitons in
fiber lasers may be considered as dissipative multi-component attractors,
rather than solitary-wave solutions of an integrable system.

The two-component states represent bright and dark vector solitons. Bright
ones are localized pulse pairs with zero background. They are typically
supported in the anomalous-GVD regime and may be understood as the direct
vector extension of scalar bright solitons. In this case, the two
polarization components are mutually trapped through XPM and may propagate
with a common group velocity even in the presence of the cavity
birefringence \cite{christodoulides1988vector}. Depending on the
birefringence and XPM strength, bright vector solitons are further
classified as \emph{group-velocity-locked} and \emph{polarization locked }
states. In the former case, the two components share a common propagation
velocity while their relative phase may evolve (is not constant) \cite%
{zhang2010vector,xu2022group}; in the latter case, the relative phase is
fixed, corresponding to a stronger form of internal coherence \cite%
{tang2005soliton,soto2000polarizationtheo,collings2000polarization,tang2008observation,zhao2009coexistence}%
. Bright vector solitons are the most commonly observed vector states in
mode-locked fiber lasers, especially in erbium- and ytterbium-doped systems
with weak or moderate birefringence \cite%
{mou2011all,lin2021vector,cui2021xpm,du2025heteronuclear}.

Dark vector solitons correspond to localized intensity dips on a
continuous-wave (CW) or quasi-CWs background and are usually associated with
normal-dispersion or defocusing-nonlinearity regimes. Their vector nature
arises when two polarization CW components carry synchronized dark notches,
or when one component supports a dark soliton while the other forms a bright
or gray companion through XPM \cite{kivshar1998dark}. Such states are
particularly interesting because they reveal the interplay between the
polarization-domain coupling and background (CW)-supported nonlinear
localization. Dark vector solitons are commonly observed in MLFLs \cite%
{zhang2010vector,ma2019observation,guo2019observation,hu2020dissipative},
and  they are theoretically interesting states as they broaden the
classification of dissipative polarization states and demonstrate that the
vector trapping is not restricted to pulses with the vanishing background.

More recently, the concept has been extended to \emph{vector SMs} and \emph{%
spatiotemporal vector states} in multimode fibers, where the birefringence,
XPM, and modal coupling jointly enrich the nonlinear dynamics \cite%
{tsatourian2013polarisation,zhao2015bound,luo2017group,wright2017spatiotemporal,luo2020stationary}%
. In the example shown in Fig. \ref{F5}(b) (borrowed from Ref. \cite%
{he2023dislocated}), a birefringence-managed fiber laser was built to create
asymmetric vector SMs, in which the cavity contains a section of a
polarization-maintaining fiber \Red{ and a polarization controller (PC) to mirror two orthogonally polarized
components}.

Overall, the bright/dark classification provides a useful theoretical
framework for understanding vector solitons in fiber lasers. Bright vector
solitons dominate in the anomalous-GVD lasers operating in the pulse regime
and are closely related to the conventional (conservative) and dissipative
pulse shaping, whereas dark vector solitons reveal how coupled nonlinear
systems can sustain localized intensity defects on top of the CW background.
Together, they highlight the essential role of internal degrees of freedom
in extending soliton physics from scalar to multi-component nonlinear wave
systems.

\subsection{PQSs (pure-quartic solitons)}

\label{subsubsec:quartic}

Beyond the classical balance between the second-order GVD and Kerr
nonlinearity, the soliton theory extends to regimes where higher-order
dispersion defines new classes of localized states. A prominent example is
PQS, These modes arise in media engineered so that the second- and
third-order GVD terms are suppressed, while FOD dominates. In this \emph{%
pure-quartic} regime, the envelope obeys an FOD-type NLSE \cite%
{blanco2016pure,de2021pure}:
\begin{equation}
\frac{\partial \psi }{\partial z}=i\frac{\beta _{4}}{4!}\,\frac{\partial
^{4}\psi }{\partial t^{4}}+i\gamma _{\mathrm{eff}}\,|\psi |^{2}\psi .
\label{E-3}
\end{equation}%
Here, the pulse evolution is governed by the interplay of FOD and SPM (Kerr
nonlinearity), yielding PQS solutions with temporal and spectral properties
distinct from those of the conventional (quadratic-GVD) solitons \cite%
{piche1996bright,roy2013formation}. In particular, PQSs exhibit a favorable
energy scaling, suggesting access to higher energies for ultrashort pulses,
which may support new soliton-laser architectures \cite{lo2018analysis}.

The first experimental observation of PQSs was reported by Blanco-Redondo
\emph{et al.} (2016). Injecting Gaussian pulses of temporal duration $1.3$
ps with the carrier wavelength $1550$~nm into a $396$~mm long
dispersion-engineered photonic-crystal waveguide \cite{li2011four} under
varying pump power, they predicted fundamental PQSs with a shape that was,
perhaps surprisingly, close to a Gaussian and demonstrated excellent
agreement with measurements. PQSs were later realized in a mode-locked fiber
laser by embedding a pulse shaper to compensate the residual GDD and impose
an effective FOD map, thereby approximating the FOD-type NLSE for the round
trip; these phases are shown in Fig. \ref{F5} \cite{runge2020pure}. In this regime, the emitted-pulse energy scales
as $E\propto T^{-3}$ with respect to the pulse duration $T$ \cite%
{tam2019stationary}, contrasting with the scaling law of the conventional
solitons, $E\propto T ^{-1}$, as per Ref. \cite{runge2020pure}. 

Subsequent studies have explored dynamical properties of PQSs, including
pulsations \cite{zhang2022pulsating,wu2025pulsating}, creeping \cite%
{han2023creeping}, and collapse/explosion regimes \cite%
{yang2024collapse,zhang2025metastable}. The PQS concept has also been
extended to bound states of PQSs \cite%
{liu2022dynamic,dai2022bound,yang2023internal,zhang2024bound}, which feature
self-similar fractal structures and bifurcations in their internal dynamics
\cite{li2024symmetry,deng2025internal}. Vector PQSs have also been
investigated in the framework of birefringent fibers \cite{zhao2021vector}
and cavities \cite{zhu2023cavity,zhu2023vector}, where the cavity
birefringence and XPM critically affect the formation and evolution \cite%
{li2025pure,wang2025surface}. Most of these results are numerical,
reflecting the challenge of simultaneously engineering FOD and nonlinear
coupling (e.g., XPM) with sufficient precision.

\subsection{Fractional solitons}

\label{subsubsec:fractional}

An emerging direction is the study of fractional solitons modeled by FNLSEs.
Before introducing the model, we aim to briefly recall its linear
counterpart, \textit{viz}., the fractional Schr\"{o}dinger equation (FSE).

The FSE was introduced in quantum mechanics via the Feynman path-integral
formalism for particles moving by \emph{L\'{e}vy flights} \cite%
{laskin2000fractional,laskin2000fractionalpath}. In this mode, the mean
displacement $l$ of the randomly walking particle from its initial position
grows with time $t$ as
\begin{equation}
l\sim t^{1/\alpha },  \label{Lwalk}
\end{equation}%
where $\alpha \leq 2$ is the \emph{L\'{e}vy index} \cite{Dubkov}. The
conventional Brownian random motion corresponds to $\alpha =2$, while $%
\alpha <2$ gives rise to faster-than-Brownian randomly-hopping mode of
motion. In the respective FSE, the kinetic-energy operator takes the form of
$\mathrm{const\cdot }\big(-\nabla ^{2}\big)^{\alpha /2}$, implying the same
length--time scaling as in Eq.~\eqref{Lwalk}. Fractional versions of
canonical quantum problems (e.g., Bohr atom and harmonic oscillator) have
been analyzed within this framework \cite{laskin2000fractals,guo2006some}.

Realizing a \emph{L\'{e}vy waveguide}(LWG) demonstrating the \emph{%
fractional diffraction} (such as in FSE), with tunable LI $\alpha $, in an
experiment is a challenge. Proposals in the realm of condensed-matter
physics were elaborated for the implementation of spatial-domain FSE in 1D
\emph{L\'{e}vy crystals}, with $1<\alpha <2$, designed as tight-binding
chains with long-range interactions \cite%
{gorenflo2001random,economou2006green,stickler2013potential}, but these
proposals have not been, as yet, realized experimentally. In optics, Longhi
\cite{longhi2015fractional} proposed to make use of the similarity between
the quantum-mechanical Schr\"{o}dinger equation and the paraxial wave
equation in a $4f$ cavity, thereby emulating the FSE as the optical paraxial
equation for light circulation in a cavity with specially tailored
transverse filtering. A temporal analogue of this scheme can be built by
mapping the required fractional temporal derivative into nonlocal
(effectively fractional) GVD in the frequency domain. Such an LWG has been
created in the experiment, enabling a mixture of the effectively fractional
and regular linear GVD with the help of a femtosecond-pulse shaper \cite%
{liu2023experimental}.

The linear operation of the optical LWG was then extended to a nonlinear
regime. In the setting, the corresponding generalized FNLSE for the slowly
varying envelope $\Psi (z,t)$ of the optical temporal signal, circulating in
the nonlinear LWG, is \cite%
{liu2023experimental,malomed2024fractional,liu2025experimental}
\begin{equation}
i\frac{\partial \Psi }{\partial z}=\left[ \frac{D}{2}\left( -\frac{\partial
^{2}}{\partial t^{2}}\right) ^{\alpha /2}+\frac{\beta _{2}}{2}\frac{\partial
^{2}}{\partial t^{2}}+i\frac{g}{2}\right] \Psi -\gamma |\Psi |^{2}\Psi ,
\label{FNLSE-LW}
\end{equation}%
where the first term on the right-hand side is the fractional time
derivative of the \emph{Riesz type}, with LI $\alpha $ and strength $D$, the
second term is the ordinary second-order GVD with coefficient $\beta _{2}$, $%
g$ is the (small-signal) gain or loss, and $\gamma $ is the Kerr
coefficient. For $g=\gamma =0$, Eq.~\eqref{FNLSE-LW} reduces to the FSE.

The Riesz derivative, which is adopted in Eq. (\ref{FNLSE-LW}), is defined
as the juxtaposition the direct and inverse Fourier transforms, thus being a
nonlocal operator in time \cite{cai2019riesz,malomed2021optical}:
\begin{equation}
\left( -\frac{\partial ^{2}}{\partial t^{2}}\right) ^{\alpha /2}\!\Psi (t)=%
\frac{1}{2\pi }\int_{-\infty }^{+\infty }\!\!|\omega |^{\alpha }\left[
\int_{-\infty }^{+\infty }e^{-i\omega (t-t^{\prime })}\Psi (t^{\prime
})\,dt^{\prime }\right] d\omega .  \label{Nonlocal}
\end{equation}%
In the frequency domain, the fractional-GVD step forward, corresponding to
Eq. (\ref{Nonlocal}), amounts to multiplication of the field by $\exp \!%
\left[ -i(Dz/2)\,|\omega |^{\alpha }\right] $. This fact enables one to
simulate Eq. (\ref{Nonlocal}) by means of the the split-step algorithm,
alternating the action of the fractional and regular GVD with the Kerr
nonlinearity, the latter contributing the familiar integral phase. The
latter phase is added to the solution, multiplying it by $\exp \!\left[
iB\,I(t)\right] $ with $B=\gamma P_{k}L_{\mathrm{n}}$ and $\,I(t)=|\Psi
|^{2}/\max (|\Psi |^{2})$, where $L_{\mathrm{n}}$ is the length of the
nonlinear segment.

Setting $g=0$, one can consider three relevant cases: (i) the regular NLSE ($%
D=0$); (ii) the pure FNLSE ($\beta _{2}=0$); and (iii) the mixed case,
including both regular and fractional GVD terms. To demonstrate the mixed
regime, a mode-locked fiber laser was constructed with a $4f$ pulse shaper
to implement the operation of the LWG. Stable \emph{fractional solitons} and
\emph{spectral-valley} states were observed in intra-cavity and extra-cavity
positions, respectively \cite{liu2023experimental,liu2025experimental}. The
inset in Fig.~\ref{F5}(e) sketches examples of fractional pulses, for LI
values $\alpha =1$ and $0.5$. The relevant operation space is conveniently
parameterized by LI $\alpha $ and the fractional-dispersion distance $L_{%
\mathrm{dis}}$ (for fixed $D$). For $L_{\mathrm{dis}}=0$, the cavity
produces conventional solitons. Increasing $L_{\mathrm{dis}}$ at fixed $%
\alpha $ can destabilize the pulse; the stability boundary corresponds to a
critical values of the fractional-dispersion length $\simeq 2T_{0}^{\alpha
}/(|D|\,\alpha )$ \cite{liu2025experimental}. In the green stability region,
the fractional solitons exhibit \emph{heavy tails} and a \emph{spectral
valley} in the temporal and frequency domains, respectively, which are
produced by the effective nonlocality representing the fractional GVD. The
extra-cavity propagation enhances the valley and can generate multi-lobe
structures, with a potential for high-dimensional encoding in optical data
transmission \cite{liu2025experimental}. Similar MLFL architectures, with
programmable fractional GVD, have been reported independently \cite%
{hoang2025nonlinear,hoang2025properties}, consistently featuring the
spectral valleys at $1\!<\!\alpha \!<\!2$.   \Red{Recent results have also demonstrated
that the relation between the pulse's width and energy can be freely adjusted, so as,
in particular, to connect solitons with the Nyquist and Hilbert shapes} \cite{hoang2026nyquist}.

Beyond these experimental advances, a broad variety of fractional-soliton
families has been analyzed theoretically in the framework of FNLSEs,
including fractional gap solitons \cite{Dong1}, DSs \cite%
{qiu2020soliton,wang2024numerical}, vortices \cite{Pengfei2}, fractional
solitons maintained by the quadratic nonlinearity \cite{quadratic},
multi-pole/multi-peak states \cite{qiu2020stabilization,zeng2021families},
clusters \cite{zeng2020preventing,li2020metastable}, and spin-orbit-coupled
settings \cite{sakaguchi2022one}. Robust breathers \cite%
{zhang2015propagation,zhang2019anomalous,tan2024manipulating}, symmetry
breaking \cite{zhang2016pts,chen2020spontaneous,ZhongSSB2023,he2024symmetry}%
, bifurcations \cite{li2021symmetry,strunin2023symmetry}, and frequency
conversion \cite{li2023second} have also been reported. For a review of the
fractional optical-wave dynamics, see Refs. \cite%
{malomed2021optical,malomed2024basic}.

\section{Bound states of solitons (alias soliton molecules, SMs)}

\label{sec:DM_double}

A remarkable manifestation of dissipative-soliton dynamics in MLFLs is the
formation of SMs (soliton molecules), in which two or more pulses are bound
by long-range interactions. The analogy to the quantum-molecular (chemical)
bonding is direct: the \emph{temporal separation} $\tau $ and the \emph{%
relative phase} $\phi $ play the roles of the bond length and angle,
respectively \cite{malomed1991bound,akhmediev1997multisoliton,cui2026bound}.
In the time domain, the SM appears as a pair of pulses separated by a
constant temporal interval $\tau $; in the frequency domain, the
interference of the bound pulses causes a spectral modulation with fringe
spacing $\propto 1/\tau $, as sketched in Fig.~\ref{F4}(a).

For a stationary SM, the relative phase is fixed in the course of the
propagation. Controlled variations of $\phi $ enable diverse dynamical
effects, including \textquotedblleft molecular vibrations\textquotedblright\
\cite{grapinet2006vibrating}, stepwise internal evolution \cite%
{herink2017real}, and phase sliding \cite{krupa2017real,wang2019optical}.
Besides $\phi $, the temporal separation (TS) $\tau $ is a fundamental
degree of freedom: it sets the interaction strength (high for $\tau $
comparable to the single-soliton width ($T$) and weak for larger
separations) and governs the nonlinear evolution, short/long-range coupling,
and the assembly of SM complexes \cite%
{malomed1991bound,Malomed1993,akhmediev1997multisoliton,gui2018soliton,wang2019optical,he2019formation,cui2026bound}%
. The TS control also plays the central role in applications, such as
multialphabetic encoding \cite{stratmann2005experimental}, optical switching
\cite{kurtz2020resonant}, and all-optical storage \cite{pang2016all},
motivating \emph{on-demand} generation and manipulation of bound states in
dissipative cavities.

\subsection{Stationary SMs}

Passively MLFLs provide an efficient platform for generating SM. Two
complementary strategies are widely used for this purpose:

\noindent \textbf{(i) Extra-cavity or pump-driven control.} Kurtz \textit{et
al.} demonstrated all-optical switching between two stationary SMs with TS
of $110$~fs and $175$~fs by means of frequency-sweeping pump modulation in a
Kerr-lens laser [Figs.~\ref{F4}(b,c)], achieving kHz-scale switching
speeds \cite{kurtz2020resonant}. While fast, this approach typically
accesses a limited set of SM states.

\noindent \textbf{(ii) Intra-cavity dispersion engineering.} Sakaguchi
\textit{et al.} showed that third-order dispersion (TOD) in the combination
with the filtering (dispersive loss) enable \emph{continuous} tuning of TS
in the SMs \cite{sakaguchi2018stationary}. In the latter model, the
effective potential
\begin{equation}
U(\tau ,\phi )=U_{0}\,e^{-i\phi -\chi \tau }\,\cos (\omega \tau )\,\cos
(\Omega \tau )  \label{E-poten}
\end{equation}%
yields two families of equilibrium TSs,
\begin{equation}
\tau _{m}^{(\omega )}=\frac{\pi }{2\omega }(1+2m),\qquad \tau _{n}^{(\Omega
)}=\frac{\pi }{2\Omega }(1+2n),  \label{E-soliton-two}
\end{equation}%
with integers $m,n$. Here $\omega \approx \delta -C\beta _{2}$ depends on
the GVD coefficient $\beta _{2}$ and filtering parameter $\delta $, and $%
\Omega \sim \beta _{3}$ depends on the TOD coefficient $\beta _{3}$.
Equations~\eqref{E-poten}-\eqref{E-soliton-two} make it explicit how GVD,
TOD, and filtering jointly set the TS values and enable continuous tuning.

Building on this approach, on-demand SM generation was demonstrated in a
hybrid MLFL that incorporates a pulse shaper to program \emph{complex GVD}
\cite{liu2022demand}:
\begin{equation}
\tilde{\beta}_{k}=\mathrm{Re}(\tilde{\beta}_{k})+i\,\mathrm{Im}(\tilde{\beta}%
_{k}),  \label{E-complex}
\end{equation}%
where $\mathrm{Re}(\tilde{\beta}_{k})$ (second-order-GVD/TOD/FOD
corresponding to $k=2,3,4,\ldots $) and $\mathrm{Im}(\tilde{\beta}_{k})$
(the dispersive loss, alias the filtering strength) are independently
controlled. In particular, the second-order loss term $\mathrm{Im}(\tilde{%
\beta}_{2})$ is \emph{necessary} to realize distinct stationary bound
states. Thus, linear TS control from $\simeq 3$ to $\simeq 5.5$~ps was
achieved, as confirmed spectrally by the modulation period in Fig.~\ref{F4}%
(e), where Fig. \ref{F4}(d) is the setup in experiment. Nearly simultaneously, a machine-learning-assisted cavity SMs with a $%
3-8$~ps TS range was achieved \cite{girardot2022demand}. Later, AI-optimized
polarization enabled TS tuning from $2$ to $58$~ps \cite{luo2024real}.

Table~\ref{tab-TS} summarizes representative experiments on TS-programmable
SMs, performed by tuning the polarization, pump power, and/or GVD. The
incorporation of AI markedly broadens the accessible TS range and improves
repeatability. For a broader context and tutorials on SM physics in MLFLs,
see Refs. \cite{gui2018soliton,mao2025temporal,singh2025review,cui2026bound}.

\begin{table}[tbph]
\caption{Experimental demonstrations of SMs (soliton molecules) with
programmable temporal separation (TS).}
\label{tab-TS}\centering
\begin{tabular}{|c|c|c|c|c|c|}
\hline
\textbf{Year} & \textbf{Configuration} & \textbf{Wavelength} & \textbf{%
Separation (ps)} & \textbf{Control parameters} & \textbf{Ref.} \\ \hline
2001 & MLFL & 1.56 $\mu$m & 1.16 & -- & \cite{Tang2001} \\ \hline
2002 & MLFL & 1.55 $\mu$m & 33.4 & -- & \cite{seong2002experimental} \\
\hline
2007 & MLFL & 1.56 $\mu$m & 3.22, 7.22, 12.50 & Pump & \cite{zhao2007bound}
\\ \hline
2011 & MLFL & 1.56 $\mu$m & 1.3, 1.8, 2.5, 9 & Pump + PC & \cite{wu2011bound}
\\ \hline
2012 & MLFL & 1.55 $\mu$m & 1.3 -- 2.2 & Pump + PC & \cite%
{yun2012generation} \\ \hline
2013 & MLFL & 1.55 $\mu$m & 4.2, 4.4, 7.1, 10.1 & Pump & \cite{Gui2013SM} \\
\hline
2015 & MLFL & 1.53 $\mu$m & 3.4 & -- & \cite{wang2015harmonic} \\ \hline
2015 & MLFL & 1.57 $\mu$m & 4.97, 7.5 & Pump + PC & \cite{wang2016black} \\
\hline
2016 & MLFL & 2 $\mu$m & 2.5 & -- & \cite{wang2016generation} \\ \hline
2017 & Kerr-lens ML & 780 nm & 0.18 & Pump modulation & \cite{herink2017real}
\\ \hline
2017 & MLFL & 1560 nm & 1.5 --  3.3 & Pump & \cite{krupa2017real} \\
\hline
2017 & MLFL & 2 $\mu$m & 7.2, 7.9 & Pump + PC & \cite{wang2016quantized} \\
\hline
2018 & MLFL & 1.56 $\mu$m & 2.6, 3.1, 8.9 & Pump + PC & \cite%
{gui2018solitonIEEE} \\ \hline
2020 & Kerr-lens ML & 0.85 $\mu$m & 0.11, 0.175 & Pump & \cite%
{kurtz2020resonant} \\ \hline
2021 & MLFL & 1.56 $\mu$m & 6.5 -- 14 & Wave plate & \cite%
{nimmesgern2021soliton} \\ \hline
2021 & MLFL & 1.55 $\mu$m & 3 -- 8 & Dispersion + AI & \cite%
{girardot2022demand} \\ \hline
2022 & MLFL & 1.56 $\mu$m & 9.1 & -- & \cite{zou2022synchronization} \\
\hline
2022 & MLFL & 1.55 $\mu$m & 3 -- 5.5 & Complex dispersion & \cite%
{liu2022demand} \\ \hline
2024 & MLFL & 1.53 $\mu$m & 4.9, 6.59, 13.14, 45.9 & Dispersion + PC & \cite%
{han2024pure} \\ \hline
2024 & MLFL & 1.55 $\mu$m & 2 -- 58 & PC + AI & \cite{luo2024real}
\\ \hline
2024 & MLFL & 1.55 $\mu$m & 1.89, 2.15 & Pump & \cite{cui2024dichromatic} \\
\hline
2025 & MLFL & 1.55 $\mu$m & 4.31, 9.52, 19.41, 42.53 & Dispersion & \cite%
{zhang2025demand} \\ \hline
\end{tabular}%
\end{table}

\subsection{Breather SMs}

\begin{figure}[!ht]
\centering
\includegraphics[width=12cm]{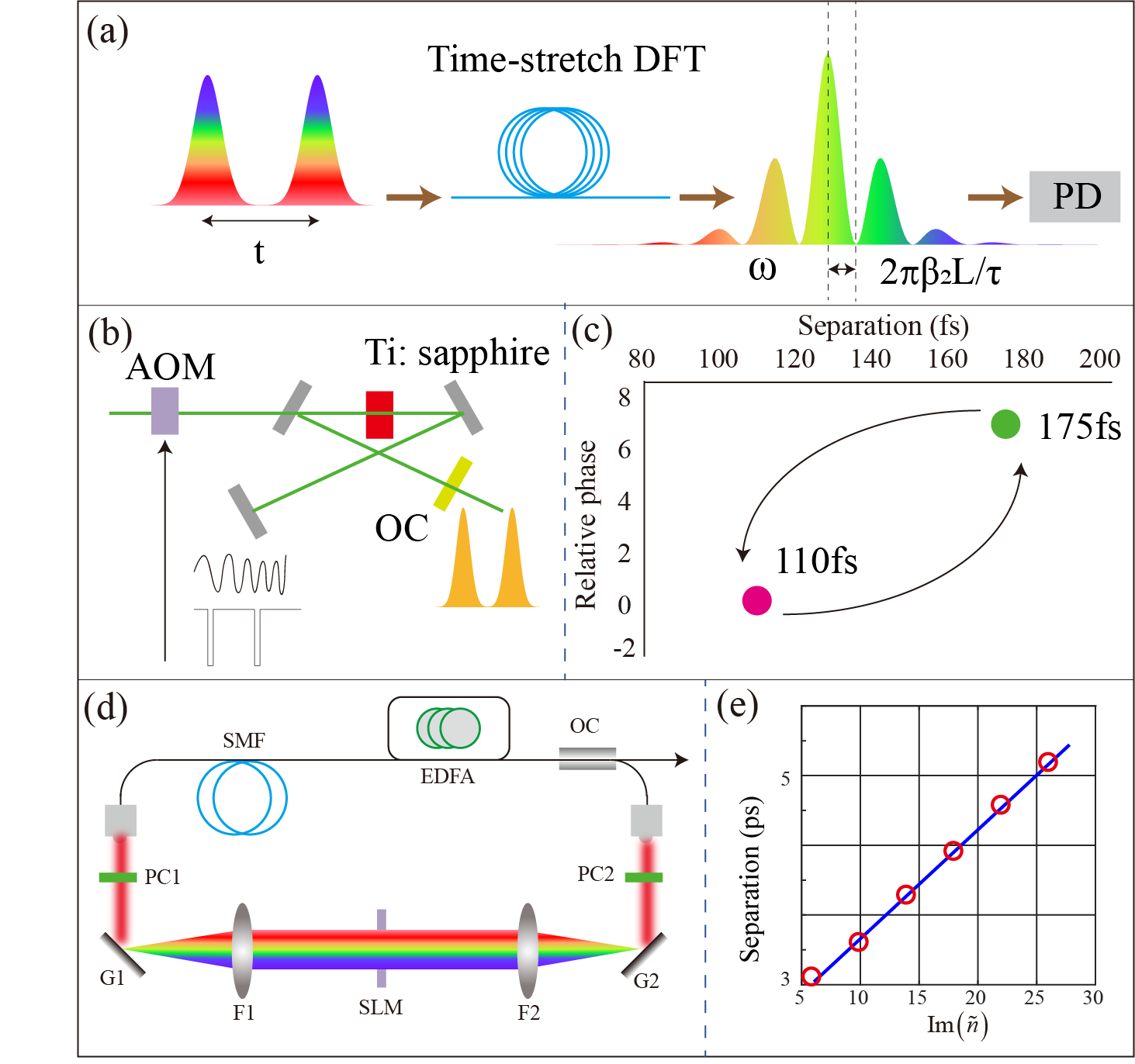}
\caption{SMs in dispersion-managed fiber-laser cavities. (a) The scheme of the time-stretched DFT and recorded single shot pulse dynamics: Two pulses in the time domain
with separation $\protect\tau $ and in the frequency domain \protect\cite%
{herink2017real,kurtz2020resonant}.
(b,c)All-optical, frequency-swept pump modulation enables switching between
stationary SMs with different separations; plots are redrawn from \protect\cite{kurtz2020resonant}.
(d,e) The intra-cavity pulse-shaper implementation of
the complex GVD for the on-demand TS control, as per \protect\cite{liu2022demand}.}
\label{F4}
\end{figure}

In addition to stationary bound states, breathing SMs constitute an
essential class of bound states in MLFLs. In this regime, the pulse's peak,
relative phase, and temporal separation oscillate periodically as functions
of the propagation distance or round-trip number, akin to breather solutions
of the NLSE, but stabilized by the cavity dissipation \cite%
{grelu2012dissipative}.

To resolve the breather evolution, single-shot diagnostics are required to
record pulse-by-pulse spectra or intensities. The time-stretched DFT
implements such measurements, using a dispersive element followed by a fast
electronic digitizer as shown in Fig. \ref{F4}(a). For an input pulse of duration $\tau $ and the
dispersion length $L\gg \tau ^{2}/(8\pi |\beta _{2}|)$, the output field is
the temporal analogue of the Fraunhofer diffraction \cite%
{bracewell1986fourier,papoulis1994pulse,berger2005complete}:
\begin{equation}
E_{\mathrm{out}}(t)\;\sim \;\exp \!\left[ i\,\frac{t^{2}}{2\beta _{2}L}%
\right] \,F\!\left( \frac{t}{\beta _{2}L}\right) ,  \label{E-TS}
\end{equation}%
where $F(\omega )=|F(\omega )|e^{i\psi (\omega )}$ is the complex spectrum, $%
\beta _{2}$ is the GVD coefficient, and $\omega $ is the detuning from the
spectral center. The detected intensity $|E_{\mathrm{out}}(t)|^{2}$ maps the
spectral intensity $I(\omega )=|F(\omega )|^{2}$ through the time-frequency
relation,
\begin{equation}
t=\beta _{2}L\,\omega .  \label{E-tw}
\end{equation}%
Expressed in terms of the wavelength, a spectral span $\Delta \lambda $ maps
to a stretched time window
\begin{equation}
\Delta \tau =|D|\,L\,\Delta \lambda ,  \label{E-deltt}
\end{equation}%
with $D$ being the fiber dispersion parameter (in units of ps$\cdot $nm$^{-1}
$\thinspace km$^{-1}$). To avoid pulse overlaps at repetition rate $R$, one
requires condition $\Delta \tau \leq R^{-1}$ \cite{goda2013dispersive}. The
spectral resolution of the time-stretched DFT, i.e., the temporal
counterpart of far-field resolution, is \cite%
{goda2009theory,goda2013dispersive,wei2017unveiling}
\begin{equation}
\delta \lambda =\lambda _{0}\,\sqrt{\frac{2}{c\,|D|\,L}},  \label{E-reso}
\end{equation}%
where $c$ is the speed of light.

Using the time-stretched DFT, Herink \emph{et al.} recorded real-time
spectral interferograms at the full laser repetition rate ($78$~MHz) over $%
\simeq 3\times 10^{5}$ consecutive round trips \cite{herink2017real}. They
directly observed the formation pathway of SMs in the MLFL: a transiently
bound pair evolving into a stationary SM. In the
phase-TS plane, trajectories spiral to a fixed point (the stationary SM).
Tuning cavity parameters (such as the pump power) reveals vibrational
dynamics, including the oscillating separation and stepwise phase evolution.

Building on the time-stretched DFT, subsequent studies have tracked diverse
manifestations of SM dynamics, including nucleation/birth \cite{liu2018real}%
, breathing \cite{krupa2017real}, data encoding \cite%
{liu2022demand,liu2023phase}, and multi-soliton pattern formation \cite%
{zhang2023heteronuclear,han2024pure,zhang2025demand}. A representative
example was done in 2017 \cite{krupa2017real}: as the pump power increases, a
single soliton first develops spectral fringes that oscillate shot-to-shot,
indicating a time-varying separation (breather SM), and then settles into a
stationary SM at a higher pump \cite{ryczkowski2018real}. Breathing features are also observed for
single-soliton states, where amplitude or energy oscillations arise from
dissipative instabilities \cite%
{ryczkowski2018real,xian2020subharmonic,zhang2024coherence}.

\section{Smart soliton lasers}

\begin{figure}[ht]
\centering
\includegraphics[width=12cm]{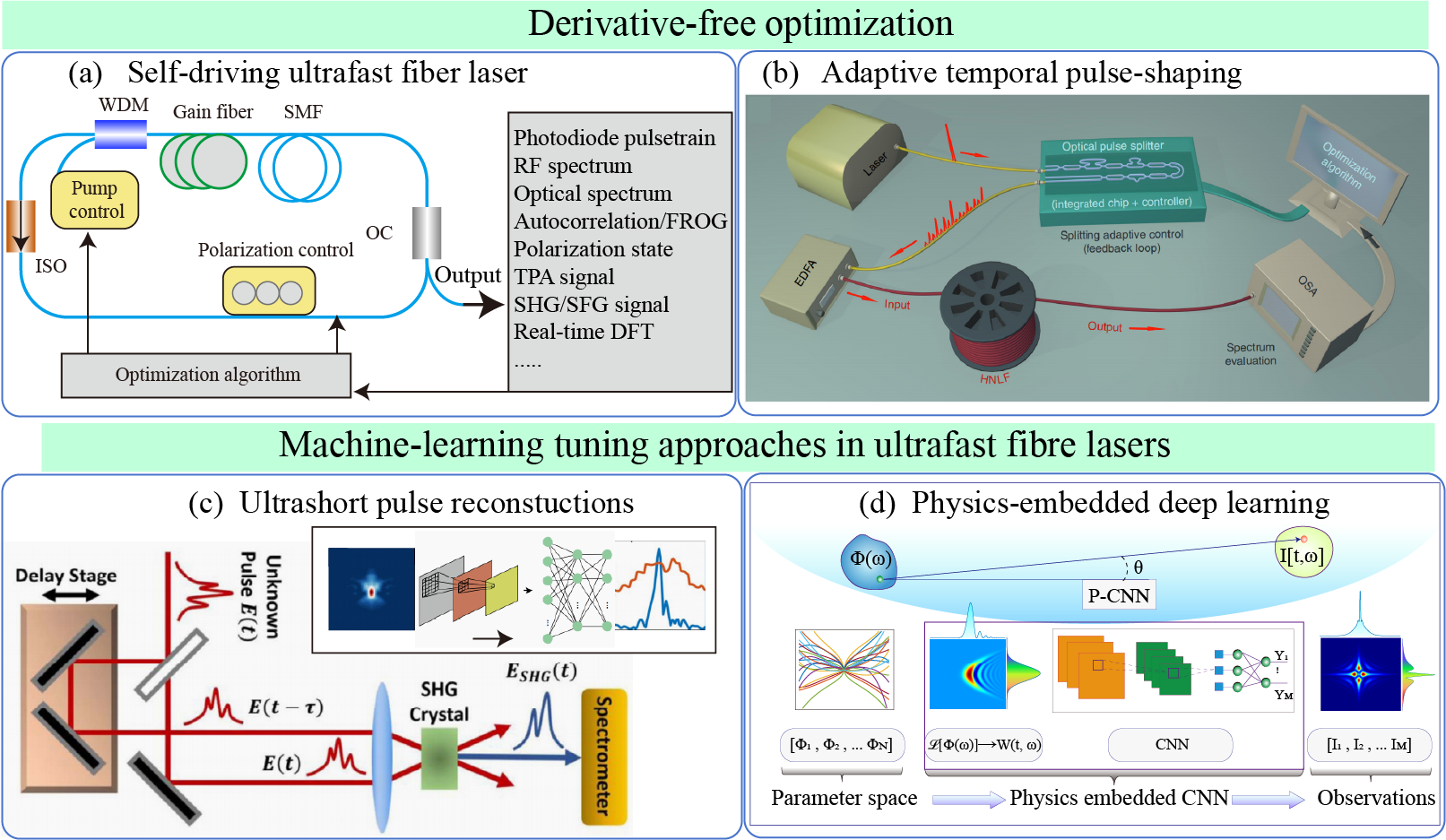}
\caption{ (a) The closed-loop control architecture for the automatic
acquisition and stabilization of target smart soliton-state regimes. The
figure is borrowed and redrawn from Ref. \protect\cite{meng2020toward}. (b) Adaptive
temporal-pulse shaping by means of GA, (borrowed from Ref. \protect\cite%
{wetzel2018customizing}, Open Access under a Creative Commons Attribution 4.0). (c) A deep neural network (DNN) was employed for
the reconstruction of ultrafast pulses (redrawn from Ref. \protect\cite%
{zahavy2018deep}, Open access). (d) A physics-embedded neural network engineering
spectro-temporal light states (redrawn from \protect\cite{liu2025engineering}, Open access). }
\label{F6}
\end{figure}


\label{sec:smater laser} Soliton operation in modern MLFLs can be cast as a
\emph{closed-loop optimization} problem: the cavity behaves as a drifting,
high-dimensional nonlinear plant, and the control objective is to discover
and maintain operating points (attractors) that deliver target pulse
attributes (energy, duration, time-bandwidth product, and long-term
stability). In this sense, the output exhibits \textquotedblleft
smart\textquotedblright\ behavior, with an external controller integrating
search/optimization strategies to adjust cavity parameters and automatically
acquire the desired state. Two typical strategies of derivative-free
optimization and machine learning are addressed in the next subsections.

\subsection{The derivative-free optimization}

\label{subsubsec:smart_soliton} As sketched in Fig.~\ref{F6}(a) \cite%
{meng2020toward}, a practical smart-laser stack comprises: (i) the cavity
(fiber or free-space), including pump, SA, gain fiber, and tunable
intracavity elements (filter, polarization, dispersion, phase/amplitude
modulation); (ii) \emph{observables}, such as the average power,
temporal/spectral intensity, RF spectrum, relative intensity noise, and
real-time diagnostics (e.g., DFT spectra); and (iii) a \emph{feedback}
stage, that compares the target and measured quality metrics and updates the
controls to reach the intended attractor.

Typical methods include genetic algorithms (GAs) and other derivative-free
optimizers. A GA initializes a population in the parameter space and
iteratively selects, recombines, and alters candidates according to a
fitness score, until the target state is reached. GAs have been used to
adjust pulse states (fundamental mode locking, CW mode locking, Q-switching)
\cite{woodward2016towards,pu2019intelligent}, optimize soliton pulse energy
with the help of tunable filters \cite{fu2013high,karar2024genetic},
generate and tailor microcombs \cite{sun2023applications,mazoukh2024genetic}%
, and program SM states \cite{gholami2014optimization,girardot2021demand}.
With programmable 4\emph{f} pulse shapers \cite{weiner2011ultrafast},
adaptive wavefront shaping has been used to optimize soliton outputs (e.g.,
broadband/SC generation \cite{lapre2023genetic,hary2023tailored}) and
multidimensional pulses in time \cite{wei2020harnessing} and spatial \cite%
{tzang2018adaptive} modes. A representative demonstration achieved direct
control of mode-locking states with output pulses as short as $40$~fs by
tailoring the cavity GVD \cite{iegorov2016direct}.

Figure \ref{F6}(b) shows a typical progress provided by an actively
controlled photonic chip to prepare and manipulate patterns of femtosecond
optical pulses in the framework of the SC generation \cite%
{wetzel2018customizing}. This approach was proposed for the customization of
nonlinear interactions via multiple pulse seeding. The preparation of
customized pulse trains was carried out with the help of an integrated
pulse-splitter \cite{park2007transform}, in which the multiple ultrashort
pulses (~with $200$ fs duration) were sent through $1$ km of a
highly-nonlinear fiber, aiming to generate a SC. Using a GA algorithm, the
system could enhance parts of the spectrum between $1300$ and $2000$ nm.

\subsection{The machine-learning (ML) control}

\label{subsubsec:smart_soliton} 
Machine-learning (ML) methods, \textit{viz}., feed-forward neural networks
(FNNs), convolutional neural networks (CNNs), and recurrent neural networks
(RNNs), provide data-driven mappings from inputs $\{X\}$ (cavity settings)
to outputs $\{Y\}$ (measured pulse metrics). Numerically, it has been shown
that parabolic, triangular, and rectangular waveforms can be generated by
propagating a conventional pulse through linear-GVD and nonlinear elements
\cite{finot2018nonlinear}.

Experimentally, ML accelerates the pulse compression to the few-cycle limit;
for example, $6.1$~fs pulses were obtained after only $42$ iterations \cite%
{farfan2018femtosecond}. Artificial neural networks have been trained to
control the pulse duration ($28-160$~ps) and spectral width ($0.75-12$~nm)
by varying the pump power \cite{kokhanovskiy2019machine}. Deep reinforcement
learning has been implemented for self-tuning of a mode-locked fiber laser
with a nonlinear loop mirror \cite{kuprikov2022deep}. A data-driven
framework combining reinforcement learning with cavity modeling has produced
target femtosecond outputs ($230-290$~fs, $16-24$~pJ) \cite{fang2023data}.
ML has also been applied to highly nonlinear soliton dynamics, such as the
SC generation. FNNs \cite%
{boscolo2020artificial,salmela2022feed,shih2023maximizing}, RNNs \cite%
{salmela2021predicting,martins2022design}, and transfer learning \cite%
{bednyakova2025transfer} predict spectral/temporal SC characteristics and
reduce the optimization time from minutes to seconds \cite%
{farfan2018femtosecond,boscolo2020artificial,hoang2022optimizing,salmela2022feed}%
. Data-driven studies have identified extreme events and classified SC
dynamics on the basis of single-shot data \cite%
{narhi2018machine,salmela2020machine,genty2021machine}.

One representative example of ML-assisted pulse characterization was
reported in 2018 \cite{zahavy2018deep}, where a DNN was employed for the
ultrafast pulse reconstruction. As shown in Fig.~\ref{F6}(c), the work used
a frequency-resolved optical gating (FROG) setup to measure the
corresponding second-harmonic--generation-FROG trace in the time--frequency
domain. By feeding these two-dimensional data into a DNN [inset in Fig.~\ref%
{F6}(c)], a network, termed FROGnet, was trained to retrieve pulse
characteristics directly from the measured trace. Although the training
process required approximately $45$~min, once trained, the network was able
to reconstruct a pulse from a single second-harmonic--generation-FROG trace
within only $0.7$~ms, demonstrating a substantial improvement in the
reconstruction speed and making the method highly attractive for real-time
experimental applications.

\subsection{The physics-involved ML (machine-learning) control}

A well-established physics-involved ML strategy relies upon the use of
physics-informed neural networks (PINNs), which represent a powerful class
of ML models designed to embed established physical laws, typically
expressed in terms of ordinary or partial differential equations, directly
into the neural-network's training process \cite%
{raissi2019physics,karniadakis2021physics}. Unlike traditional purely
data-driven neural networks, whose performance is highly dependent on the
quantity and quality of data, PINNs leverage the known physical constraints
to produce an additional loss term that quantifies the residual of the
governing physical equations, thereby enhancing the generality and
robustness of the ML scenarios, especially those with limited or noisy data.

An early application showing the potential for predicting complex nonlinear
dynamics, including the SC generation, was reported by J. M. Dudley's group
\cite{salmela2021predicting}, using neural networks (a foundation for the
PINN approach). Then, more work was done in this direction, to study the
pulse dynamics in fibers \cite{jiang2022physics}, predict various optical
soliton states (bright, dark, SMs) \cite{fang2022predicting,wu2022prediction}%
, and compare the performance with that provided by traditional methods \cite%
{wang2022applications,jiang2023predicting}.

A physics-embedded neural network is another effective physics-involved ML
strategy \cite{ren2024deep,xu2024physics}, as shown in Fig. \ref{F6}(d), in
which the physics constraints or transform are added to the front of the
network. By combining with CNN, the system forms a physics-embedded
convolution neural network (P-CNN), which can be connected to the parameter
space $\Phi (\omega )$ and target space $I[t,\omega ]$ -- in particular, to
learn the pulse evolution in highly nonlinear fibers for SC engineering \cite%
{liu2025engineering}. Instead of purely data-driven features, this technique
involved the Wigner function, which represents the field in the chronocyclic
domain \cite{walmsley2009characterization,najafabadi2024intensity},\
\begin{equation}
W(t,\omega )=\!\int_{-\infty }^{+\infty }\!\!A\!\left( t+\tfrac{t^{\prime }}{%
2}\right) A^{\ast }\!\left( t-\tfrac{t^{\prime }}{2}\right) e^{i\omega
t^{\prime }}\,dt^{\prime },  \label{E-Wigner}
\end{equation}%
was used to address spectrotemporal correlations.

This representation helps CNNs to suppress uncorrelated noise and reduce
sensitivity to phase variations of the carrier envelope \cite{zuo2022deep}.
In simulations, the trained P-CNN achieved higher accuracy (with average
regression $R\approx 0.97$) and threefold lower loss than an FNN ($R\approx
0.91$). Experimentally, two P-CNNs enabled continuous tuning of the SC
spectral/temporal features (the central frequency, bandwidth, pulse
duration) and directly produced $12$~fs pulses (sub-three-cycle ones) from a
fiber SC source, without external compression \cite%
{dudley2006supercontinuum,piccoli2021intense}.

Concise reviews on ML and nonlinear photonics are available in Refs. \cite%
{genty2021machine,ma2023artificial}.

\section{Conclusion and perspectives}

\label{sec:conclusion}

This chapter provides a review of the vast topic of the nonlinear
soliton-wave control in modern fiber systems, from Hamiltonian to
dissipative and generalized dissipative operation regimes. Particular
emphasis was placed on MLFLs (mode-locked fiber lasers) as programmable
testbeds that support a wide family of states, including DSs (dissipative
solitons) with desirable energy and duration, DM (dispersion-managed)
single-pulse states (similaritons, vector solitons, PQS (pure-quartic
solitons), and fractional solitons), and double-pulse bound states, i.e.,
stationary and breather SM (soliton molecules). Real-time diagnostics (DFT
(dispersive Fourier transform) and time-lens), together with the algorithmic
control (derivative-free optimization and ML (machine learning)) integrate
the theory, device design, and robust operation for smart soliton lasers.
These capabilities enable advanced pulse synthesis and point toward
soliton-based ultrafast sources for data processing and analog photonic
computing.

\paragraph{Outlook.}

Several directions appear especially promising for controlling solitons in
nonlinear fiber-based and hybrid platforms:

\begin{itemize}
\item \textbf{Real-time metrology and closed-loop manipulation.} The
combination of the DFT/time-lens instruments with the phase retrieval to
develop the single-shot spectroscopy of transient events (fission,
explosions, SM breathing) and use these signals as the feedback to stabilize
or steer dynamics on the round-trip basis.

\item \textbf{High-resolution spectral-temporal shaping.} Overcoming\ the
bandwidth and resolution limits of intracavity elements (gain media,
filters, pulse shapers) by co-designing broadband, low-loss programmable
dispersion terms (including higher-order and fractional ones) to realize
finer spectral and temporal control.

\item \textbf{Multi-pulse nonlinear dynamics.} Systematic studies of
interactions between multiple pulses, mediated by XPM, Raman-induced forces,
dispersive-wave binding, vector/spin degrees of freedom, and mapping
stability/transition diagrams for complex soliton assemblies.

\item \textbf{Computing with fiber-optic solitons.} The development\ of
soliton-based neuromorphic and analog processors, including reservoir
computing, optical Ising machines, and photonic memories/logic, leveraging
programmable cavities and the learning-in-the-loop control.

\item \textbf{Integration and hybridization.} Translation of fiber-based
schemes to chip-scale platforms (e.g., Kerr microcombs with on-chip
dispersion engineering), and design of hybrid fiber-chip cavities with
embedded sensors/actuators and machine-learning controllers for the creation
of compact resilient soliton engines.
\end{itemize}


\end{document}